\documentclass[conference]{IEEEtran}
\usepackage[T1]{fontenc}
\usepackage{microtype}
\usepackage{graphicx}
\usepackage{amssymb}
\usepackage[cmex10]{amsmath}
\usepackage{flushend}
\usepackage{color}
\begin{document}

\newcommand{\todo}[1]{\textbf{[\textcolor{red}{TODO:} \textcolor{blue}{#1}]}}

\newcommand{\hy}{\text{-}}
\newcommand{\un}{\text{\_}}

\newcommand{\mf}[1]{\mathfrak{#1}}
\newcommand{\ms}[1]{\mathsf{#1}}
\newcommand{\mc}[1]{\mathcal{#1}}
\newcommand{\mi}[1]{\mathit{#1}}

\newtheorem{lemma}{Lemma}
\newtheorem{theorem}{Theorem}
\newtheorem{example}{Example}
\newtheorem{convention}{Convention}
\newtheorem{remark}{Remark}

\newcommand{\inference}[2]{\begin{array}{r@{}l} {\cal I} & \strut\displaystyle {#1} \over \strut\displaystyle {#2}\\ \end{array}}

\title{Automated Verification of Interactive Rule-Based Configuration Systems (Additional Material)}

\author{
  \IEEEauthorblockN{Deepak Dhungana\IEEEauthorrefmark{1}, Ching Hoo
    Tang\IEEEauthorrefmark{2}, Christoph
    Weidenbach\IEEEauthorrefmark{2}, Patrick Wischnewski
    \IEEEauthorrefmark{3}}
  \IEEEauthorblockA{\IEEEauthorrefmark{1}Siemens AG
    \"Osterreich\\ Vienna, Austria\\ Email:
    deepak.dhungana@siemens.com}
  \IEEEauthorblockA{\IEEEauthorrefmark{2}Max-Planck-Institute for
    Informatics\\ Saarbr\"ucken, Germany\\ Email: \{chtang,
    weidenbach\}@mpi-inf.mpg.de}
  \IEEEauthorblockA{\IEEEauthorrefmark{3}Logic4Business
    GmbH\\ Saarbr\"ucken, Germany\\ Email:
    patrick.wischnewski@logic4business.com}
 }

\onecolumn

\copyright 2013 IEEE. Personal use of this material is
permitted. Permission from IEEE must be obtained for all other uses,
in any current or future media, including reprinting/republishing this
material for advertising or promotional purposes, creating new
collective works, for resale or redistribution to servers or lists, or
reuse of any copyrighted component of this work in other works.

This is a version of a paper accepted to the 2013 28th IEEE/ACM International 
Conference on Automated Software Engineering (ASE) 
(DOI: 10.1109/ASE.2013.6693112) 
that contains additional 
material in section~\ref{sec:pidl} and \ref{sec:model-dopler}.

\twocolumn

\maketitle

\begin{abstract}
Rule-based specifications of systems have again
become common in the context of product line variability modeling and
configuration systems. In this paper, we define a logical
foundation for rule-based specifications that has enough expressivity 
and operational behavior to be
practically useful and at the same time enables decidability of
important overall properties such as consistency or cycle-freeness. 
Our logic supports rule-based interactive user transitions as well as 
the definition of a domain theory via rule transitions.
As a running example, we model DOPLER, a rule-based configuration system
currently in use at Siemens.
\end{abstract}

\section{Introduction}\label{sec:introduction}

After their first successful application in the context of expert
systems in the 1980's, rule-based specifications of systems have again
become common in the context of product line variability modeling and
configuration systems.  Designing a rule-based language is always a
compromise between expressivity, semantics, and decidability of
overall properties. Expressivity starts at simple logics and ranges up
to full programming language availability. Semantics, meaning what is
the result of applying a set of rules in a particular context, ranges
from a programming language style operational semantics to a model
theoretic semantics. Finally, depending on the expressivity and
underlying semantics, proving properties of a rule-based language
ranges from polynomial decidability to undecidability.

In this paper, we investigate the role of rule-based languages in the
context of interactive product configuration. 
Interactive configuration refers to the process of a user
interactively assigning values to variables, under given restrictions
specified using rules. Each step in the user-configurator interaction
includes a user selecting a value from a domain and the configurator
executing applicable rules to propagate the change. Our goal is to
define a logical
foundation that has enough expressivity and operational behavior to be
practically useful and at the same time enables decidability of
important overall properties such as consistency. 

We will start from the available language
DOPLER~\cite{DhunganaGR11}, which is a product line variability
modeling tool-set currently in use at Siemens. A first attempt towards a 
formal semantics for DOPLER has been
previously discussed in a workshop paper~\cite{DhunganaHR10}. The
initial workshop paper describes the key concepts of DOPLER, however a 
more comprehensive semantic framework that can eventually be 
subject to an automated analysis of existing knowledge bases is still
missing. This paper provides a model-theoretic semantics for
interactive rule-based systems, in particular DOPLER.

Our new logic PIDL (Propositional Interactive Dynamic Logic, see Section~\ref{sec:pidl})
serves as a framework for the modeling, analysis and execution of rule-based
configuration systems.
It supports three fundamentally different types of formulas. The first formula type 
are constraints. Constraints describe necessary conditions of any configuration, e.g.,
that two components can never go together. The second type are rule transitions.
Rule transitions describe necessary changes to the configuration typically as a
result of a user decision, e.g., a user has selected two components but for technical
reasons they have to be replaced by a third, different component.
Finally, the third formula type are user transitions. They describe changes to the
configuration done by a user in an interactive way, e.g., she selects a certain
component. The semantics of these formula types is inherently different.
Constraints must always be fulfilled while rule and user transitions must not lead to 
an inconsistent state including an appropriate notion of update.
In PIDL, a user transition is only applicable if the exhaustive application of rule
transitions reaches a unique consistent state, called rule-terminal
state. The latter condition distinguishes PIDL from
any other framework for describing rule-based systems, like guarded transition systems
or temporal logics that lack language constructs supporting our semantics of rule and
user transitions. It is in particular this semantics that enables a deep analysis of rule
and user transitions including properties like confluence or cyclicity (see Section~\ref{sec:properties}).
When analyzing a PIDL specification user transitions are considered in a non-deterministic,
exhaustive way.

In Section~\ref{dec:DOPLER}, we present examples of PIDL properties which can be effectively
analyzed and which are often highly indicative for errors in a rule-based configuration
knowledge base. They include inconsistency (conflicting rules, constraints), incompleteness
(missing rules), redundancy (redundant rules), circularity
(circularly depending rules), and confluence (result unique rule-based computations). The presented
framework is field-tested, and has proved to be adequate to
detect these errors in existing models. A summary of the results is
presented in Section~\ref{sec:model-dopler}.

Our main contribution is the new logic PIDL (see
Section~\ref{sec:pidl}) motivated by the semantics of DOPLER (see
Section~\ref{dec:DOPLER}). The logic is expressive enough to model
DOPLER and at the same time it offers decidability of important
properties of rule-based systems, such as inconsistency,
incompleteness, redundancy, confluence, and cyclicity. 
This way it generalizes known  approaches such as guarded transitions
systems. At the same time it replaces the problem of undecidability of programming language
verification applicable to rule-based systems written in a programming language by
an expensive, but effective decision procedure for all the above properties.
In particular, PIDL is expressive enough to support decision revision as expressed 
by rules of the form $A \land \phi \leadsto \{ \neg A,\ldots\}$ and the concept of rule-terminal
states. We show by a first prototypical
implementation that PIDL can in fact be turned into a useful software
system for the practical analysis of rule-based systems.

\section{Illustrative Example: DOPLER}\label{dec:DOPLER}

\begin{figure*}[tbh!]
\begin{center} 
  \includegraphics [width=500pt]{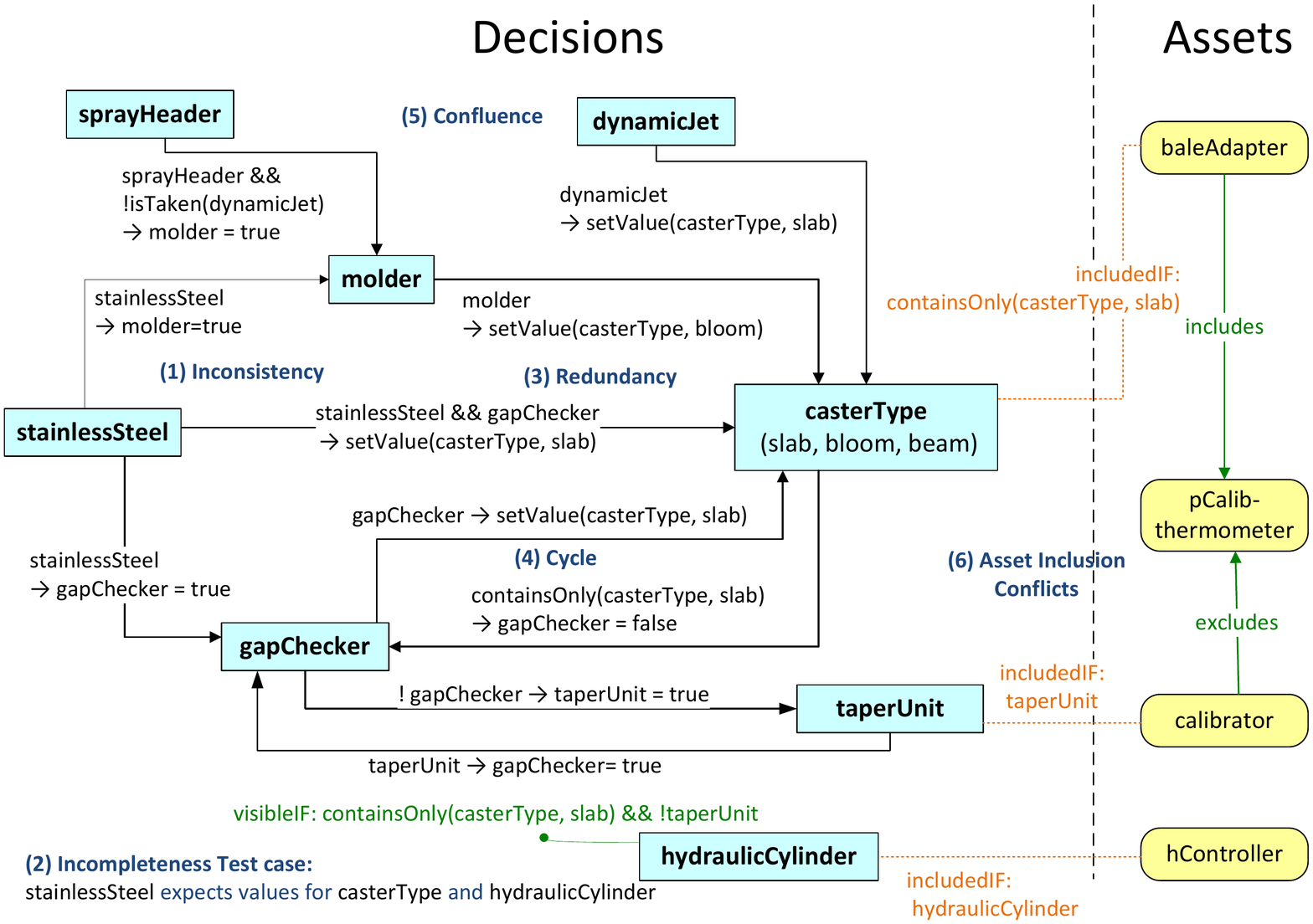}
\end{center}
\caption{Illustration of a DOPLER Model depicting decisions,
  assets and dependencies among them (rules, visibility
  conditions and asset inclusion conditions).}%
\label{fig:dopler-example}
\end{figure*}

DOPLER is a rule-based tool suite for interactive product
configuration.  A DOPLER model describes the differences between
products in a product line. The key modeling elements are
\emph{decisions} (representing configuration variables) and
\emph{assets} (representing artifacts being configured).  Dependencies
among decisions are modeled using rules of the form \texttt{if
  <condition> then <action>}. The assets are associated with decisions
through boolean expressions called inclusion conditions. Assets may
``include'' or ``exclude'' other assets. Further details on the
modeling approach have been described previously \cite{DhunganaGR11}. 
A DOPLER model
serves as the running example in this paper. We present an example
from the steel plant automation domain.
Figure~\ref{fig:dopler-example} depicts the key modeling elements
and dependencies among them.

DOPLER models are used for interactive configuration. During
configuration, a set of decisions is taken by the user. Some other
decisions are assigned appropriate values through rules, which get
executed after each user interaction. Each decision has an associated
visibility condition to specify whether the variable is currently
accessible to the user.

The operational semantics of DOPLER models can be informally described
as follows. At runtime, decisions can either be visible or invisible
to the user. All visible decisions (visibility condition evaluates to
\texttt{true}) are presented to the user and the user assigns a value. Every
user interaction triggers the rule engine, which evaluates all the
rules and executes them if they are applicable. Rule execution can
cause a variable binding, which leads to a recursive call of the rule
engine. The user can also change the values of already taken
decisions. Changing an already taken decision also causes a roll-back
of the previous rule execution caused by the same decision. This
ensures that the effect of the rules is undone when the condition of
a rule no longer holds. An asset can either be included in or excluded
from the desired final product (evaluation of the inclusion condition
of the assets or the evaluation of an asset dependency). A state in
DOPLER (the current assignment of values to decisions) can therefore
be changed by user interactions and the subsequent execution of rules.

The running example (cf.~Figure~\ref{fig:dopler-example}) shows decisions 
and assets as well as the relationships between them. Decisions are depicted
by rectangle boxes on the left part of the image and assets by rounded corner boxes on
the right part of the image. An arrow leading from
a decision A to another decision B indicates that changing the value of A may
also have an effect on B, depending on whether the rule, written as a label
\texttt{<condition>} $\rightarrow$ \texttt{<action>} of the arrow, gets 
activated. If \texttt{<condition>} evaluates to \texttt{true}, the action is 
executed.
The condition parts are written as usual Java-style Boolean expressions. The 
DOPLER framework also provides functions to manipulate the values of decisions 
and to query decision
values, such as \texttt{setValue} and \texttt{isTaken}. For instance,  
\texttt{dynamicJet} evaluating to \texttt{true} makes \texttt{casterType} 
have the value
\texttt{slab}. The former is a \emph{Boolean} decision which can be assigned 
the value \texttt{true} or \texttt{false}, the latter is an \emph{enumeration}
decision whose range of possible finitely many values is predefined.

In this example, three
decisions are visible to the user from the beginning:
\texttt{sprayHeader}, \texttt{dynamicJet} and
\texttt{stainlessSteel}. 
The decision \texttt{hydraulicCylinder} has a visibility condition, namely,
it requires  \texttt{casterType}'s only value to be \texttt{slab}  
and \texttt{taperUnit} to be \texttt{false}.
If the visibility condition is evaluated to \texttt{true} then this decision is visible as well.
The rest of the decisions are not visible and thus cannot be taken directly
by the user. 

The lines connecting assets and
decisions represent the assets' inclusions conditions. Their evaluation depend
on the decisions and determine whether assets get included in the product or
not. The asset \texttt{hController} is 
included if its inclusion condition \texttt{hydraulicCylinder} evaluates 
to \texttt{true}. \texttt{baleAdapter} is included similarly and additionally
requires the other asset \texttt{pCalibthermometer}. This is one example
of the inclusion/exclusion relationships between assets.

The running example presents examples of different kinds of anomalies
in a DOPLER Model.
\begin{itemize}
\item \textbf{Inconsistency} occurs when the execution of
  different rules in the rule base leads to conflicting values
  for a decision. For example, assignment of the decision
  \texttt{stainlessSteel = true} would result in \texttt{casterType =
    bloom} through the rules associated with \texttt{molder}
  and at the same time \texttt{casterType = slab} through the
  rules associated with \texttt{gapChecker}. This is an
  anomaly in the model, as the decision \texttt{casterType}
  can have only one value.
\item \textbf{Incompleteness} occurs when an expected configuration
  cannot be reached due to the lack of transitions. For
  example, the modeler expects the value of
  \texttt{hydraulicCylinder} to be set automatically,
  after \texttt{stainlessSteel} is assigned a
  value. However, there is no rule that would lead to
  this state.
\item \textbf{Redundancy} occurs when more than one rule is
  modeled to achieve the same effect in the rule base. For
  example, after the value of \texttt{stainlessSteel} is set,
  we have \texttt{casterType = slab} through two different
  paths. This is an anomaly because it increases the
  maintenance effort of the rule base.
\item \textbf{Cyclicity} occurs when the propagation of rules never stops
  because the involved rules change the variables such that there is
  always another rule that can be executed. For example, the three
  variables \texttt{gapChecker}, \texttt{taperUnit} and
  \texttt{casterType} form a cycle. The variable \texttt{gapChecker}
  is changed by \texttt{casterType} and \texttt{taperUnit}, making a
  different rule applicable after each execution.
\item \textbf{Violation of Confluence} occurs when the order in which
  decisions are taken has an impact on the final configuration
  result. For example, depending on whether \texttt{sprayHeader} or
  \texttt{dynamicJet} is assigned a value first, the value of the
  variable \texttt{casterType} is either \texttt{bloom} or
  \texttt{slab}.
\item \textbf{Asset Inclusion Conflicts} occur
  when the inclusion conditions of the assets
  are not consistent with the dependencies
  among the assets. For example, when
  \texttt{baleAdapter} and \texttt{calibrator} are both
  included in the configuration, they have a
  conflicting dependency to
  \texttt{pCalibthermometer} and it is not
  clear whether \texttt{pCalibthermometer}
  should be included or excluded.
\end{itemize}

\section{PIDL: Propositional Interactive Dynamic Logic} \label{sec:pidl}

PIDL is a new logic that provides detailed models for configuration
systems. In particular, and in addition to all other temporal or
dynamic proposition logics, it provides the notion of a rule-terminal
state.  Rule-terminal states are normal forms or fixed points with
respect to a subset of the transition rules. They will later on be
used to distinguish rules caused by user decisions from rules
describing the domain. For the latter we expect uniqueness of the
description, i.e., any user decision
leads to a unique new state with respect to the domain rules.

We first describe the syntax and semantics of PIDL and then provide
a sound and complete calculus for it, based on the ideas of 
superposition~\cite{BachmairGanzinger01handbook,NieuwenhuisRubio01handbook,Weidenbach01handbook}.
This calculus constitutes a decision procedure that will then be used
in the rest of paper to actually analyze the properties of rule-based
systems, in particular DOPLER.

Let $F_{\Pi}$ denote the set of all propositional formulas over 
a finite set of propositional variables $\Pi$. A 
\emph{state} is a consistent set of literals from $\Pi$.

A \emph{PIDL specification} $\mf{S}$  is a 5-tuple 
$(\Pi, S_I, \ms{C}, T_U, T_R)$, where
\begin{itemize}
\item $\Pi$ is a finite set of propositional variables,
\item $S_I$ is the \emph{initial} state,
\item $\mathsf{C}$ is a finite set of propositional formulas over $\Pi$ called 
\emph{constraints},
\item $T_U$ is a finite set of indexed tuples $\chi_i \leadsto E_i$ called 
\emph{user transitions}, where $\chi_i\in F_{\Pi}$ and $E_i$ is a state,
\item $T_R$ is a finite set of indexed tuples $\chi_j \leadsto E_j$ called 
\emph{rule transitions}, where $\chi_j\in F_{\Pi}$ and $E_j$ is a state,
\end{itemize}
and we assume that all user and rule transitions have different indexes.
The set $\Pi$ contains a dedicated variable $\mi{start}$ that is not used elsewhere
in the specification.

Starting from the initial state $S_I$, the specification $\mf{S}$
induces a number of states.  An \emph{update} of a state $S$ by an
$E$, written $S \triangleleft E$, is defined by $S\triangleleft E :=
\{L | (L \in S \text{ and } \overline{L} \not \in E) \text{ or } L \in
E\}$. Literals in $S$ are replaced by
the literals in $E$ that are of the same variable but have a different
sign, and literals of $E$ previously not contained in $S$ are added to
$S$.
\begin{example}
  $\{A, B, \neg C\} \triangleleft \{\neg B, \neg C, D\} = 
  \{A, \neg B, \neg C, D\}$ 
\end{example}
A \emph{rule transition application} using a rule transition $\chi_i
\leadsto E_i \in T_R$ is a tuple $S \rightarrow_i S'$, where 
\begin{itemize}
\item $S \cup \ms{C} \not \models \bot$,
\item $S \cup \ms{C} \models \chi_i$, and
\item $S' = S \triangleleft E_i$.
\end{itemize}

\begin{example}
The state $S = \{A, \neg B, C\}$ induces via rule transition
  $A \land C \leadsto  \{ B\}$ the state $S \triangleleft E_i = \{A, B, C\}$.
\end{example}

For convenience, we may use the term rule transition instead of
rule transition application.

We say that a state $S$ is \emph{rule-terminal} if 
for all $\chi_i \leadsto E_i \in T_R : S \cup \ms{C}
\models \chi_i$  implies $(S = S \triangleleft E_i)$. A state $S$ is
therefore rule-terminal if no rule transition leads to a state that is
different from $S$. 

\begin{example}
The state $S = \{A, B, \neg C, D, \neg E\}$ is rule-terminal with
respect to the set of rule transitions $T_R$ that consists of the
following transitions:
\begin{itemize}
\item $A \leadsto \{B, \neg C\}$,
\item $\neg C \leadsto \{D\}$, and
\item $A \land \neg D \leadsto \{E\}$.
\end{itemize}
\end{example}

A \emph{user transition application} using a user transition $\chi_i
\leadsto E_i$ is a tuple $S \rightarrow_i S'$, where
\begin{itemize}
\item $S \cup \ms{C} \not \models \bot$,
\item $S$ is rule-terminal,
\item $S \cup \ms{C} \models \chi_i$, and
\item $S' = S \triangleleft E_i$.
\end{itemize}

The conditions are the same as for rule transitions expect we have an
additional requirement that the state $S$ must be rule-terminal. As in the 
case of rule transitions, we may use the term user transition for user 
transition application. Given a configuration system, all possible
user interactions are modeled by user transitions in PIDL.

A \emph{path} $\tau$ from state $S_1$ to $S_n$ is a finite list of 
indexes $[i_1, i_2, \dots, i_{n-1}]$, 
such that $S_j\rightarrow_{i_j} S_{j+1}$.
The empty path  is
denoted by $[]$. In other words, a path $\tau$ consists of
indexes that correspond to the user and rule transitions. We want to
be able to construct paths incrementally in our calculus. To this end,
we use the notation $[i_1, i_2, \dots, i_n] :: i \; := \; [i_1, i_2,
  \dots, i_n, i]$ to denote the extension of paths. Furthermore, the
\emph{length of a path} $\tau$ is denoted by $|\tau|$ and is the
number of elements it contains.

The set of all states that are reachable from the initial state $S_I$
is denoted by $\mc{S}_{\mf{S}}$: \[\mc{S}_{\mf{S}} := \{S | \text{
  there is a path from } S_I \text{ to } S\}.\] 
Note that $\mc{S}_{\mf{S}}$ is well-defined, i.e., all
$S \in \mc{S}_{\mf{S}}$ are consistent sets of literals, i.e., they
do not contain any complementary literals. The reason for this is that
the initial state $S_I$ is consistent by definition, and the update
operations that define the states of $\mc{S}_{\mf{S}}$ preserve this
property.

An \emph{interpretation $\mc{I}$} of a specification $\mc{S}_{\mf{S}}$
is a function \[\mc{I} : \mc{S}_{\mf{S}} \rightarrow 2^{\Pi}\] such
that $\mc{I}(S)\models S$ and $\mi{start}\in\mc{I}(S)$ for all $S\in
\mc{S}_{\mf{S}}$.

The interpretation of a single state yields a Herbrand interpretation, so
$\mc{I}(S)\models A$ if $A\in\mc{I}(S)$ and $\mc{I}(S)\models \neg A$
if $A\not\in\mc{I}(S)$.

An interpretation $\mc{I}$ is a \emph{model} of a specification
$\mf{S}$ if $\mc{I}(S) \models \ms{C}$
for all $S \in \mc{S}_{\mf{S}}$. A specification is called
\emph{inconsistent} if it has no model.

\begin{example}
Assume the following specification $\mf{S} = (\Pi, S_I, \ms{C}, T_U,
T_R)$ defined by
\begin{itemize}
\item $\Pi = \{A, B, C, D\}$,
\item $S_I = \{\neg A, \neg B\}$,
\item $\ms{C} = \{B \rightarrow C\}$,
\item $T_U = \{\neg A \leadsto \{A, B\}\}$ and
\item $T_R = \{C \leadsto \{D\}\}$.
\end{itemize}
Then $\mc{S}_{\mf{S}}$ consists of the following states:
\begin{itemize}
\item $S_I = \{\neg A, \neg B\}$,
\item $S_1 = \{A, B\}$,
\item $S_2 = \{A, B, D\}$.
\end{itemize}
One possible interpretation $\mc{I}$ is
\begin{itemize}
\item $\mc{I}(S_I) = \emptyset$,
\item $\mc{I}(S_1) = \{A, B, C\}$, and
\item $\mc{I}(S_2) = \{A, B, C, D\}$.
\end{itemize}
Another interpretation $\mc{I'}$ is
\begin{itemize}
\item $\mc{I'}(S_I) = \emptyset$,
\item $\mc{I'}(S_1) = \{A, B\}$, and
\item $\mc{I'}(S_2) = \{A, B, D\}$.
\end{itemize}
$\mc{I}$ is a model of $\mf{S}$, whereas $\mc{I'}$ is not.
\end{example}

The calculus for PIDL
operates on clauses annotated with labels which are
representative of the states induced by the specification. We show how 
those clauses are generated and what inference steps can be applied to them.

A \emph{labeled clause} has the form $(S, \tau, p \, || \, C)$, where
$S$ is a state, $\tau$ is a path, $p \in \mathbb{N} \cup \{*\}$, and
$C$ is a propositional clause over $\Pi$ including the
variable $\mi{start}$.

One important concept is the ordering of labeled clauses which plays a
role for redundancy and in proving completeness of the calculus. The
ordering on clauses is based on a total ordering on paths. We define
$\tau \prec \tau'$ if
\begin{itemize}
\item $|\tau| < |\tau'|$, or
\item $|\tau| = |\tau'|$ and $\tau <_{\mi{lex}} \tau'$,
\end{itemize}
where $<_{\mi{lex}}$ is the lexicographic extension of the
$<$-ordering on natural numbers.

\begin{example}
\begin{itemize}
\item Let $\tau_1 = [3, 5, 2]$ and $\tau_2 = [4, 9, 2, 1, 3]$. Then
  $\tau_1 \prec \tau_2$ because $|\tau_1| = 3 < 5 = |\tau_2|$.
\item Let $\tau_3 = [2, 9, 4, 2]$ and $\tau_4 = [2, 9, 4, 5]$. Then
  $\tau_3 \prec \tau_4$ because $|\tau_3| = 4 = |\tau_4|$ and $\tau_3
  <_{\mi{lex}} \tau_4$.
\end{itemize}
\end{example}

Intuitively, we associate a propositional clause $C$ with the state it
is derived from. What derived means is made more precise by the calculus
description below. In addition to the state itself, the label of a
clause also contains the path, i.e., the sequence of rule
applications which led to this state. Furthermore, the symbol $p$
indicates whether $C$ is a general clause of the state, in which case
$p = *$, or a specific clause connected to a rule condition, in which
case $p \in \mathbb{N}$. The special $\mi{start}$ clause functions as
the ``first clause'' of each state in the calculus.

An interpretation $\mc{I}$ for a specification $\mf{S}$ is a
\emph{model} of a labeled clause $(S, \tau, * \, || \, C)$, 
written $\mc{I} \models (S, \tau, * \, || \, C)$, if
$S \in \mc{S}_{\mf{S}}$ and $\mc{I}(S) \models C$. Moreover, $\mc{I}$
is a model of a set of labeled clauses if $\mc{I}$ is a model of each
clause of the set.
In the rest of the paper, we may refer to labeled clauses simply as
clauses if the context is clear.

As usual as for a superposition-based calculus, redundancy and model
assumptions are defined with respect to a total ordering lifted from
the propositional variables to clauses.

Let $\prec$ be a total ordering on $\Pi$. It can be lifted to literals
by $P\prec \neg P\prec Q$ if $P\prec Q$. Then it is lifted to clauses
by its multiset extension on literals and finally to a partial
ordering on labeled clauses by
  $$(S, \tau, p \, || \, C) \prec (S', \tau', p' \, || \, C')$$ if 
 $[\tau \prec \tau']$ or $[\tau = \tau'$, $p = *$ or $p = p'$,
  and $C \prec C']$.  Note that $\prec$ is well-founded on labeled
clauses.

\begin{example}
Let $S, S'$ be two states and $\tau = [3, 1, 6], \tau' = [8, 2, 6, 1]$
be two paths. Furthermore, let the following ordering on variables be
given: $A \prec B$.
\begin{itemize}
\item $(S, \tau, * \, || \, A \lor B) \prec (S', \tau', * \, || \,
  \neg B)$ because $\tau \prec \tau'$.
\item $(S, \tau, * \, || \, A \lor B \prec (S', \tau, * \, || \, \neg
  B))$ because $A \lor B \prec \neg B$.
\end{itemize}
\end{example}

A labeled clause $(S, \tau, p \, || \, C)$ is \emph{redundant} with
respect to a set $N$ of labeled clauses if there are clauses $(S,
\tau', p_1' \, || \, C_1)$, $(S, \tau', p_2' \, || \, C_2)$, $\dots$,
$(S, \tau', p_n' \, || \, C_n) \in N$ with $(S, \tau', p_i' \, || \,
C_i) \prec (S, \tau, p \, || \, C)$ for $1 \leq i \leq n$ and $C_1,
C_2, \dots, C_n \models C$.

\begin{example}
\begin{itemize}
\item $(S, [5, 6, 9], * \, || \, A \lor B)$ is redundant with respect
  to $\{(S, [2, 3], * \, || \, A)\}$ because \[(S, [2, 3], * \, || \,
  A) \prec (S, [5, 6, 9], * \, || \, A \lor B)\] and $A \models A \lor
  B$.
\item $(S, [5, 6, 9], * \, || \, A \lor B)$ is redundant with respect
  to $\{(S, [5, 6, 9], * \, || \, B)\}$ because \[(S, [5, 6, 9], * \,
  || \, B) \prec (S, [5, 6, 9], * \, || \, A \lor B)\] and $B \models
  A \lor B$.
\end{itemize}
\end{example}

Our notion of redundancy prevents the duplication of clauses sharing the same
state at all: in the presence of a clause $(S, \tau, * \, || \, \mi{start})$
any other clause $(S, \tau', * \, || \, \mi{start})$ with $\tau\prec\tau'$
is redundant.

We now describe the inference rules $\mi{SInf}$ consisting of
Units Creation, Constraints Creation, User Transition Condition Creation,
Rule Transition Condition Creation,
 Factoring, and Superposition that serve as
a calculus with respect to the
specification $\mf{S} = (\Pi, S_I, \ms{C}, T_U, T_R)$ for reasoning
in one particular state. 

\begin{itemize}
\item \textbf{Units Creation:} $\inference{S, \tau, * \, || \,
  \mi{start}}{S, \tau, * \, || \, L}$,\\[0.2cm] where $L \in S$.
\item \textbf{Constraints Creation:} $\inference{S, \tau, * \, || \,
  \mi{start}}{S, \tau, * \, || \, C}$,\\[0.2cm] where $C \in
  \mi{cnf}(\ms{C})$.
\end{itemize}
Units and Constraints Creation take the start clause $(S, \tau, * \,
|| \, \mi{start})$ of the state and produce labeled clauses
for the  constraints $\ms{C}$ and unit clauses out the state literals.
Each literal of the state and each constraint is
represented as labeled clauses by virtue of the two rules. It will
become apparent below where the start clause comes
from. $\mi{cnf}(N)$ is the set of clauses that is the result of
transforming a set of propositional formulas $N$ into conjunctive normal form.
\begin{itemize}
\item \textbf{User Transition Conditions Creation:}\\[0.2cm]
  $\inference{S, \tau, * \, || \, \mi{start}}{S, \tau, i \, || \,
  C}$,\\[0.2cm] where $C \in \mi{cnf}(\neg \chi_i), \chi_i \leadsto
  E_i \in T_U$.
\item \textbf{Rule Transition Conditions Creation:}\\[0.2cm]
  $\inference{S, \tau, * \, || \, \mi{start}}{S, \tau, i \, || \,
  C}$,\\[0.2cm] where $C \in \mi{cnf}(\neg \chi_i), \chi_i \leadsto
  E_i \in T_R$.
\end{itemize}
Having the start clause as premise, these rules yield labeled clauses
that represent the conditions $\chi_i$ of the user and rule
transitions. The propositional clauses $C$ come from the negated
conditions $\chi_i$ because we want to work with refutations, which
will be explained more precisely. The label of such a clause contains
the index $i$ of the transition it corresponds to.
\begin{itemize}
\item \textbf{Factoring:} $\inference{S, \tau, p \, || \, C \lor A
  \lor A}{S, \tau, p \, || \, C \lor A}$,\\[0.2cm] where $C$ is a
  propositional clause and $A$ is a literal.
\item \textbf{Superposition:}\\[0.2cm] $\inference{S, \tau, p \, || \,
  C \lor L \qquad S, \tau, p' \, || \, D \lor \overline{L}}{S, \tau, p
  \oplus p' \, || \, C \lor D}$,\\[0.2cm] where
    \begin{itemize}
    \item $L$ and $\overline{L}$ are maximal in their respective clauses with
      respect to $\prec$,
    \item $p = *$ or $p' = *$ or $p = p'$, and
    \item the value of $p \oplus p'$ is defined by
      \begin{equation*}
        p \oplus p' =
        \begin{cases}
          p' & \text{, if } p = *,\\ p & \text{, if } p' = * \text{ or
          } p = p'
        \end{cases} \quad .
      \end{equation*}
    \end{itemize}
\end{itemize}
The two rules largely resemble rules of the well-known resolution
calculus \cite{BachmairGanzinger01handbook}. Factoring produces clauses where
duplicate literals are removed. Superposition is resolution on the
labeled clauses where the $p$ in the label of the conclusion clause
indicates if the result is connected to the transitions ($p = i$) or
not ($p = *$).

Given a set of inference rules, such as $\mi{SInf}$, we define
$N^0_{\mi{SInf}} = N$, $N^{i+1}_{\mi{SInf}} = N^i\cup \{(S, \tau, p \,
|| \, C) | (S, \tau, p \, || \, C)$ is a conclusion of an $\mi{SInf}$
inference with premises in $N^i\}$, and $N^*_{\mi{SInf}} := \bigcup_{i
  \geq 0} N^i_{\mi{SInf}}$.

Now the inference rules $\mi{SRInf}$ include the rules $\mi{SInf}$
plus the rules Forward Rule Transition and Forward User Transition
defined below.

\begin{itemize}
\item \textbf{Forward Rule Transition:}\\[0.2cm] $\inference{S, \tau,
  i \, || \, \bot}{S', \tau :: i, * \, || \, \mi{start}}$,\\[0.2cm]
  where
  \begin{itemize}
  \item $(S, \tau, * \, || \, \bot) \not \in \{(S, \tau, * \, || \,
    \mi{start})\}^*_{\mi{SInf}}$
  \item $\chi_i \leadsto E_i \in T_R$,
  \item $S' = S \triangleleft E_i$.
  \end{itemize}
\item \textbf{Forward User Transition:}\\[0.2cm] $\inference{(S, \tau,
  i \, || \, \bot)}{S', \tau :: i, * \, || \, \mi{start}}$,\\[0.2cm]
  where
  \begin{itemize}
  \item $S = S \triangleleft E_j$ for each $(S, \tau, j \, || \, \bot)
    \in \{(S, \tau, * \, || \, \mi{start})\}^*_{\mi{SInf}}$
  \item $(S, \tau, * \, || \, \bot) \not \in \{(S, \tau, * \, || \,
    \mi{start})\}^*_{\mi{SInf}}$
  \item $\chi_i \leadsto E_i \in T_U$,
  \item $\chi_j \leadsto E_j \in T_R$, and
  \item $S' = S \triangleleft E_i$.
  \end{itemize}
\end{itemize}
Forward Rule Transition says that whenever there is a clause $(S,
\tau, i \, || \, \bot)$ corresponding to a rule transition $\chi_i
\leadsto E_i$, and the inferences $\mi{SInf}_{\mf{S}}$ starting with
$(S, \tau, * \, || \, \mi{start})$ have not yielded $(S, \tau, * \, ||
\, \bot)$, we can derive a new start clause $(S', \tau :: i, * \, ||
\, \mi{start})$ with $S'$ being the state $S$ updated by the rule
transition and the transition being stored in the path $\tau$. Forward
User Transition works analogously with the additional premise that for
all $(S, \tau, j \, || \, \bot)$ corresponding to rule transitions
derived , the updates of the current state $S$ by the rule transitions
does not change the state.

\begin{remark}\label{rem:start-clauses}
We observe that each derivation of a clause labeled with $S$ and
$\tau$ must start with $(S, \tau, * \, || \, \mi{start})$ except for
the starting clause itself which is derived through the transition
rules.
\end{remark}

\medskip
\begin{lemma}[Local Soundness of $\mi{SRInf}$]\label{lem:local-soundness}
Let $\mf{S} =(\Pi, S_I, \ms{C}, T_U, T_R)$ be a PIDL specification.
Then
\begin{enumerate} 
\item $(S, \tau, * \, || \, C) \in \{(S_I, [], * \, || \,
\mi{start})\}^*_{\mi{SRInf}} \Rightarrow S \cup \ms{C} \models C$ and
\item $(S, \tau, i \, || \, C) \in \{(S_I, [], * \, || \,
\mi{start})\}^*_{\mi{SRInf}}
    \Rightarrow S \cup \ms{C} \cup \{\neg \chi_i\} \models
    C$,\\[0.2cm] where $C \neq \mi{start}$ and $\chi_i \leadsto E_i
    \in T_U \cup T_R$.
\end{enumerate}
\end{lemma}

\medskip
\begin{IEEEproof}
\begin{enumerate}
\item By induction on the derivation length of $(S, \tau, * \, || \,
    C)$ relative to $(S, \tau, * \, || \, \mi{start})$.
    \begin{itemize}
    \item Base case:  By remark \ref{rem:start-clauses}, $(S, \tau, * \, || \,
      C)$ is the conclusion of a units creation or constraints
      creation inference, with $(S, \tau, [] \, || \,
      \mi{start})$ being the premise. Then $C = L$ with $L \in S$ or
      $C \in \mi{cnf}(\ms{C})$ respectively. In both cases, $S \cup
      \ms{C} \models C$.
    \item Induction step:
      \begin{itemize}
        \item $(S, \tau, * \, || \, C)$ is the conclusion of a
          factoring inference, with a clause $(S, \tau, * \, || \, C'
          \lor A \lor A)$ being the premise and $C = C' \lor A$. By
          induction hypothesis, $S \cup \ms{C} \models C' \lor A \lor
          A$. Then surely, $S \cup \ms{C} \models C' \lor A$.
        \item $(S, \tau, * \, || \, C)$ is the conclusion of a
          superposition inference with clauses $(S, \tau, * \, || \,
          C_1 \lor L)$ and $(S, \tau, * \, || \, C_2 \lor
          \overline{L})$ being the premises and $C = C_1 \lor C_2$. By
          induction hypothesis, $S \cup \ms{C} \models C_1 \lor L$ and
          $S \cup \ms{C} \models C_2 \lor \overline{L}$. By soundness
          of propositional resolution, it then holds that $S \cup
          \ms{C} \models C_1 \lor C_2$.
      \end{itemize}
    \end{itemize}
  \item By induction on the derivation length of $(S, \tau, i \, || \,
    C)$ relative to $(S, \tau, * \, || \, \mi{start})$.
    \begin{itemize}
      \item Base case: The base case is defined by 
        remark~\ref{rem:start-clauses}. Then, $(S, \tau, i \, || \, C)$ is the
        conclusion of a user or a rule transition conditions creation,
        with $(S, \tau, \epsilon \, || \, \mi{start})$ being the
        premise. In both cases, $C \in \mi{cnf}(\neg \chi_i)$ and thus
        $S \cup \ms{C} \cup \{\neg \chi_i\} \models C$.
      \item Induction step:
        \begin{itemize}
          \item $(S, \tau, i \, || \, C)$ is the conclusion of a
            factoring inference, with a clause $(S, \tau, i \, || \,
            C' \lor A \lor A)$ being the premise and $C = C' \lor
            A$. By induction hypothesis, $S \cup \ms{C} \cup \{\neg
            \chi_i\} \models C' \lor A \lor A$. Then surely, $S \cup
            \ms{C} \cup \{\neg \chi_i\} \models C' \lor A$.
          \item $(S, \tau, i \, || \, C)$ is the conclusion of a
            superposition inference with clauses $(S, \tau, p_1 \, ||
            \, C_1 \lor L)$ and $(S, \tau, p_2 \, || \, C_2 \lor
            \overline{L})$ being the premises and $C = C_1 \lor
            C_2$. It holds that $p_1 = i$ or $p_2 = i$. Without loss
            of generality, let $p_1 = i$. By induction hypothesis, $S
            \cup \ms{C} \cup \{\neg \chi_i\} \models C_1 \lor L$. Now
            we consider two possibilities for $p_2$: If $p_2 = i$,
            then also by induction hypothesis $S \cup \ms{C} \cup
            \{\neg \chi_i\} \models C_2 \lor \overline{L}$. If $p_2 =
            *$, then we have $S \cup \ms{C} \models C_2 \lor
            \overline{L}$ as shown above, and therefore $S \cup \ms{C}
            \cup \{\neg \chi_i\} \models C_2 \lor \overline{L}$. In
            any case, $S \cup \ms{C} \cup \{\neg \chi_i\} \models C_1
            \lor C_2$ by soundness of propositional resolution.
        \end{itemize}
    \end{itemize}
\end{enumerate}
\end{IEEEproof}

\medskip
\begin{lemma}[Local completeness of $\mi{SRInf}$]\label{lem:local-completeness}
Let $\mf{S} =(\Pi, S_I, \ms{C}, T_U, T_R)$ be a PIDL specification.
Furthermore, let $S \in \mc{S}_{\mf{S}}$ and $(S, \tau, * \, || \,
\mi{start}) \in \{(S_I, [], * \, || \,
\mi{start})\}^*_{\mi{SRInf}}$. Then 
\begin{enumerate}
\item $S \cup \ms{C} \models \bot
\Rightarrow (S, \tau, * \, || \, \bot) \in \{(S_I, [], * \, || \,
\mi{start})\}^*_{\mi{SRInf}}$ and
\item $S \cup \ms{C} \cup \{\neg \chi_i\} \models \bot \Rightarrow
    (S, \tau, i \, || \, \bot) \in \{(S_I, [], * \, || \,
\mi{start})\}^*_{\mi{SRInf}}$.
\end{enumerate}
\end{lemma}

\medskip
\begin{IEEEproof}
\begin{enumerate}
\item Since $(S, \tau, * \, || \, \mi{start}) \in \{(S_I, [], * \, || \,
\mi{start})\}^*_{\mi{SRInf}}$, we can derive labeled clauses annotated
by $S$, $\tau$ and $*$ by units creation and constraints creation. The
assumption $S \cup \ms{C} \models \bot$ together with the definition
of the superposition inference rule and refutational completeness of
propositional resolution means indeed $(S, \tau, * \, ||
\, \bot) \in \{(S_I, [], * \, || \, \mi{start})\}^*_{\mi{SRInf}}$.
\item Analogously to 1). In addition to the labeled clauses inferred by units
creation and constraints creation, we can derive labeled clauses annotated by
$S$, $\tau$ and $i$ by user transition conditions creation and rule transition
conditions creation. The assumption $S \cup \ms{C} \cup \{\neg \chi_i\} 
\models \bot$ together with the definition of the superposition inference
rule and refutational completeness of propositional resolution means 
indeed $(S, \tau, i \, || \, \bot) \in \{(S_I, [], * \, || \,
\mi{start})\}^*_{\mi{SRInf}}$
\end{enumerate}
\end{IEEEproof}

\medskip
\begin{theorem}[Soundness and Completeness of $\mi{SRInf}$]
Let $\mf{S} =(\Pi, S_I, \ms{C}, T_U, T_R)$ be a PIDL specification.
Then $\mf{S}$ is inconsistent iff there is a labeled clause $(S, \tau,
* \, || \, \bot)\in \{(S_I, [], * \, || \,
\mi{start})\}^*_{\mi{SRInf}}$.
\end{theorem}

\medskip
\begin{IEEEproof}
  $(\Rightarrow)$ Let $\mf{S}$ be inconsistent, i.e., there is a state
  $S \in \mc{S}_{\mf{S}}$ such that $S \cup \ms{C} \models \bot$.

We first prove that there is a labeled clause $(S, \tau, * \,
|| \mi{start}) \in \{(S_I, [], * \,
|| \, \mi{start})\}^*_{\mi{SRInf}}$. We do this by induction on the
length $n$ of a fixed path from $S_I$ to $S$.

$n = 0$: The path consists of just the initial state $S_I$, and indeed
$(S_I, [], * \, || \, \mi{start}) \in
\{(S_I, [], * \, || \, \mi{start})\}^*_{\mi{SRInf}}$.

$n \rightarrow n + 1$: The path has the form $S_I \rightarrow S_1
\rightarrow \dots \rightarrow S_n \rightarrow_i S$. By induction
hypothesis, $(S_n, \tau_n, * \, || \, \mi{start}) \in
\{(S_I, [], * \, || \, \mi{start})\}^*_{\mi{SRInf}}$. By the definition of paths,
there are two cases:
\begin{itemize}
\item $S_n \rightarrow_i S$ is a rule transition, i.e., $\chi_i
  \leadsto E_i \in T_R$. By definition,
  \begin{enumerate}
  \item $S_n \cup \ms{C} \not \models \bot$,
  \item $S_n \cup \ms{C} \models \chi_i$,
  \item $S = S_n \triangleleft E_i$.
  \end{enumerate}
  Then
  \begin{enumerate}
  \item $\Rightarrow (S_n, \tau_n, * \, || \, \bot) \not \in
    \{(S_I, [], * \, || \, \mi{start})\}^*_{\mi{SRInf}}$ by lemma
    $\ref{lem:local-soundness}$, and
  \item $\Rightarrow S_n \cup \ms{C} \cup \{\neg \chi_i\} \models
    \bot$ \\ $\Rightarrow (S_n, \tau_n, i \, || \, \bot) \in
    \{(S_I, [], * \, || \, \mi{start})\}^*_{\mi{SRInf}}$ (lemma 
    \ref{lem:local-completeness}).
  \end{enumerate}
  We have the correct premises and forward rule transition of
  $\mi{SRInf}$, yielding a clause $(S, \tau, * \, || \,
  \mi{start})$ with $\tau = \tau_n :: i$.
\item $S_n \rightarrow_i S$ is a user transition, i.e., $\chi_i
  \leadsto E_i \in T_U$. By definition,
  \begin{enumerate}
  \item $S_n \cup \ms{C} \not \models \bot$,
  \item $S_n$ is rule-terminal,
  \item $S_n \cup \ms{C} \models \chi_i$,
  \item $S := S_n \triangleleft E_i$.
  \end{enumerate}
  Then
  \begin{enumerate}
  \item $\Rightarrow (S_n, \tau_n, * \, || \, \bot) \not \in \{(S_I,
    [], * \, || \, \mi{start})\}^*_{\mi{SRInf}}$ by lemma
    $\ref{lem:local-soundness}$,
  \item $\Leftrightarrow (S_n \cup \ms{C} \models \chi_i \Rightarrow
    S_n = S_n \triangleleft E_i$ for each $\chi_i \leadsto E_i \in
    T_R)$\\ $\Rightarrow (S_n \cup \ms{C} \cup \{\neg \chi_i\} \models
    \bot \Rightarrow S_n = S_n \triangleleft E_i$ for each $\chi_i
    \leadsto E_i \in T_R)$\\ $\Rightarrow ((S_n, \tau_n, i \, || \,
    \bot) \in \{(S_I, [], * \, || \, \mi{start})\}^*_{\mi{SRInf}}
    \Rightarrow S_n = S_n \triangleleft E_i$ for each $\chi_i \leadsto
    E_i \in T_R)$ (lemma \ref{lem:local-completeness})\\ $\Rightarrow
    ((S_n, \tau_n, i \, || \, \bot) \in (\{(S_n, \tau_n, * \, || \,
    \mi{start})\})^*_{\mi{SInf}} \Rightarrow S_n = S_n \triangleleft
    E_i$ for each $\chi_i \leadsto E_i \in T_R)$ 
    (remark \ref{rem:start-clauses}), and
  \item $\Rightarrow S_n \cup \ms{C} \cup \{\neg \chi_i\} \models
    \bot$ \\ $\Rightarrow (S_n, \tau_n, i \, || \, \bot) \in \{(S_I,
        [], * \, || \, \mi{start})\}^*_{\mi{SRInf}}$ (lemma
        \ref{lem:local-completeness}).\\
  \end{enumerate}
\end{itemize}
  so we can apply forward user transition of $\mi{SRInf}$ and
  we get a clause $(S, \tau, * \, || \, \mi{start})$ with $\tau =
  \tau_n :: i$.

  We conclude that there is indeed a labeled clause $(S, \tau, * \,
  || \mi{start}) \in \{(S_I, [], * \,
  || \, \mi{start})\}^*_{\mi{SRInf}}$. By the assumption and
  lemma \ref{lem:local-completeness}, we have that there is a labeled
  clause $(S, \tau, * \, || \, \bot)\in \{(S_I, [], * \,
  || \, \mi{start})\}^*_{\mi{SRInf}}$.

\medskip
 $(\Leftarrow)$ Assume there is a labeled clause 
$(S, \tau, * \, || \, \bot)\in 
\{(S_I, [], * \, || \, \mi{start})\}^*_{\mi{SRInf}}$. By remark \ref{rem:start-clauses}, there is a labeled clause $(S, \tau, * \, || \, \mi{start})\in 
\{(S_I, [], * \, || \, \mi{start})\}^*_{\mi{SRInf}}$.

We first prove that $S \in \mc{S}_{\mf{S}}$ whenever there is a
labeled clause $(S, \tau, * \, || \, \mi{start})\in \{(S_I, [], * \,
|| \, \mi{start})\}^*_{\mi{SRInf}}$. We do this by induction on the
path $\tau$.

Let $\tau = []$. Then $(S_I, [], * \, || \, \mi{start})
\in \{(S_I, [], * \, || \, \mi{start})\}^*_{\mi{SRInf}}$ by the initialization 
rule, and indeed $S_I \in \mc{S}_{\mf{S}}$.

Let $\tau = \tau'::i$. We have a labeled clause $(S, \tau' :: i, * \,
|| \, \mi{start}) \in \{(S_I, [], * \, || \, \mi{start})\}^*_{\mi{SRInf}}$. 
According to
the rules of $\mi{SRInf}$, there are two cases in which this
clause could have been derived:
\begin{enumerate}
\item By forward rule transition.\\[0.2cm] $\inference{S', \tau', i \,
  || \, \bot}{S,
  \tau' :: i, * \, || \, \mi{start}}$,\\[0.2cm] 
  where
  \begin{itemize}
  \item $(S', \tau', * \, || \, \bot) \not \in \{(S', \tau', * \, || \, \mi{start})\}^*_{\mi{SInf}}$,
  \item $\chi_i \leadsto E_i \in T_R$, and
  \item $S = S' \triangleleft E_i$.
  \end{itemize}
  From the premise we conclude $(S', \tau', i \, || \, \bot) \in
  \{(S_I, [], * \, || \, \mi{start})\}^*_{\mi{SRInf}}$. As observed in remark
  \ref{rem:start-clauses}, $(S', \tau', * \, || \, start) \in
  \{(S_I, [], * \, || \, \mi{start})\}^*_{\mi{SRInf}}$. By induction 
  hypothesis, $S' \in
  \mc{S}_{\mf{S}}$. We show that in $S'$ we can apply a rule
  transition with $\chi_i \leadsto E_i$ as defined.
  \begin{itemize}
  \item $(S', \tau', * \, || \, \bot) \not \in 
    \{(S', \tau', * \, || \, \mi{start})\}^*_{\mi{SInf}}$
    first means $(S', \tau', * \, || \, \bot) \not \in
    \{(S_I, [], * \, || \, \mi{start})\}^*_{\mi{SRInf}}$, 
    since $(S', \tau', * \, || \, \bot)$  could only be derived by 
    $\mi{SInf}$ inferences starting from $(S', \tau', * \, || \, start)$,  
    and we already know $(S', \tau', * \, || \, start)
    \in \{(S_I, [], * \, || \, \mi{start})\}^*_{\mi{SRInf}}$ from
    above. By
    lemma \ref{lem:local-soundness}, $S' \cup \ms{C} \not \models \bot$.
  \item As observed above, $(S', \tau', i \, || \, \bot) \in
    \{(S_I, [], * \, || \, \mi{start})\}^*_{\mi{SRInf}}$. 
      By lemma \ref{lem:local-soundness},
      $S' \cup \ms{C} \cup \{\neg \chi_i\} \models \bot$ and thus $S' \cup
      \ms{C} \models \chi_i$.
    \item Finally, $S = S' \triangleleft E_i$ indeed as required by the
      inference rule.
    \end{itemize}
  The conditions of a rule transition as defined are fulfilled, so 
  we can do a rule transition
  $S' \rightarrow_i S$ using $\chi_i \leadsto E_i$ and thus $S \in
  \mc{S}_{\mf{S}}$.
\item By forward user transition.\\[0.2cm]
  $\inference{S', \tau', i \, || \, \bot}
  {S, \tau' :: i, * \, || \, \mi{start}}$,\\[0.2cm]
  where
  \begin{itemize}
  \item  $S' = S' \triangleleft E_j$ for each $(S', \tau, j \, || \, \bot) \in
    \{(S', \tau', * \, || \, \mi{start})\}^*_{\mi{SInf}}$
  \item $(S', \tau, * \, || \, \bot) \not \in \{(S', \tau, * \, || \, \mi{start})\}^*_{\mi{SInf}}$
  \item $\chi_i \leadsto E_i \in T_U$,
  \item $\chi_j \leadsto E_j \in T_R$, and
  \item $S = S' \triangleleft E_i$.
  \end{itemize}
  The premise $(S', \tau', i \, || \, \bot)$
  indicates that $(S', \tau', i \, || \, \mi{start}) \in
  \{(S_I, [], * \, || \, \mi{start})\}^*_{\mi{SRInf}}$. By induction hypothesis, 
  $S' \in \mc{S}_{\mf{S}}$. We show that in $S'$ we can apply a user transition
  with $\chi_i \leadsto E_i$.
  \begin{itemize}
  \item The premise $(S', \tau', * \, || \, \bot) \not \in $
    entails $S' \cup \ms{C} \not \models \bot$, which can be shown
    analogously to the forward rule transition case.
\item From the premise
[$S' = S' \triangleleft E_j \text{ for each }$ \\
  $(S', \tau, j \, || \, \bot) \in
    \{(S', \tau', * \, || \, \mi{start})\}^*_{\mi{SInf}}$] we get that each time 
we have
  \begin{itemize}
  \item $(S', \tau, j \, || \, \bot) \in
    \{(S', \tau', * \, || \, \mi{start})\}^*_{\mi{SInf}}$, which means
  \item $(S', \tau', j \, || \, \bot) \in
    \{(S_I, [], * \, || \, \mi{start})\}^*_{\mi{SRInf}}$ analogously to 1), which
     means
  \item $S' \cup \ms{C} \cup \{\neg \chi_j\} \models \bot$ by lemma
    \ref{lem:local-soundness}, which means
  \item $S' \cup \ms{C} \models \chi_j$,
  \end{itemize}
  it holds that $S' = S' \triangleleft E_j$ for all $\chi_j \leadsto
  E_j$, so $S'$ is rule-terminal.
\item As in case 1), the premise $(S', \tau', i \, || \, \bot)$ gives
  us $S' \cup \ms{C} \models \chi_i$.
\item $S = S' \triangleleft E_i$ indeed as required by the inference
  rule.
\end{itemize}
The requirements of a user transition are satisfied and we can apply a user
transition $S' \rightarrow_i S$ by $\chi_i \leadsto E_i$, and thus $S
\in \mc{S}_{\mf{S}}$.
\end{enumerate}
Now we have $S \in \mc{S}_{\mf{S}}$, and there is a labeled clause $(S,
\tau, * \, || \, \bot)\in \{(S_I, [], * \, || \,
\mi{start})\}^*_{\mi{SRInf}}$ by assumption. By lemma \ref{lem:local-soundness},
it holds that $S \cup \ms{C} \models \bot$, so $\mf{S}$ is
inconsistent.
\end{IEEEproof}

\medskip
\begin{theorem}[Decidability of PIDL] \label{theo:decidability}
Let $\mf{S} =(\Pi, S_I, \ms{C}, T_U, T_R)$ be a PIDL specification.
Then $\{(S_I, [], * \, || \, \mi{start})\}^*_{\mi{SRInf}}$ is finite
up to redundancy.
\end{theorem}

\medskip
\begin{IEEEproof}
Let $N$ be such a subset of $\{(S_I, [], * \,
|| \, \mi{start})\}^*_{\mi{SRInf}}$ that for each $(S, \tau, * \,
|| \, \mi{start}) \in N$ there is no $\tau'$ with $\tau' \prec \tau$
and $(S, \tau', * \, || \, \mi{start}) \in \{(S_I, [], * \,
|| \, \mi{start})\}^*_{\mi{SRInf}}$. $N$ is not empty, because $(S_I,
[], * \, || \, \mi{start}) \in N$ and the ordering $\prec$ on paths is
well-founded. $N$ is also finite because $\{S \mid (S, \tau, p \,
|| \, C) \in \{(S_I, [], * \, || \, \mi{start})\}^*_{\mi{SRInf}}\}$ is
finite ($S$ is a set of literals over the finite variable set $\Pi$)
and the path ordering is total. Let $M$ be another subset of $\{(S_I,
[], * \, || \, \mi{start})\}^*_{\mi{SRInf}}$ such that $M
:= \bigcup_{(S, \tau, * \, || \, \mi{start}) \in N} \{(S, \tau, * \,
|| \, \mi{start})\}^*_{\mi{SInf}}$.

Now note that $\{(S, \tau, * \, || \, \mi{start})\}^*_{\mi{SInf}}$ is
finite up to redundancy because it corresponds to propositional
resolution on clauses labeled with $S$ and $\tau$ which is known to be
finite up to redundancy. Consequently, $M$ is also finite up to
redundancy. Let $(S, \tau, p \, || \, C) \in \{(S_I, [], * \,
|| \, \mi{start})\}^*_{\mi{SRInf}}$ and $(S, \tau, p \, || \,
C) \not \in M$. By assumption, there is a $\tau' \prec \tau$ with
$(S, \tau', * \, || \, \mi{start}) \in \{(S_I, [], * \,
|| \, \mi{start})\}^*_{\mi{SRInf}}$, which means there is a labeled
clause $(S, \tau', p \, || \, C) \in \{(S_I, [], * \,
|| \, \mi{start})\}^*_{\mi{SRInf}}$ making $(S, \tau, p \, || \, C)$
redundant with respect to $M$.
\end{IEEEproof}

\medskip
It is well-known that $\mi{SInf}$ terminates on propositional logic with
respect to redundancy, corresponding here to reasoning on clauses sharing
the same path and state label.

Exploring Theorem~\ref{theo:decidability}, given the saturation 
$N^* = \{(S_I, [], * \, || \, \mi{start})\}^*_{\mi{SRInf}}$ of a PIDL
specification $\mf{S}$, the \emph{state graph} $G_\mf{S}$ of $\mf{S}$ consists 
of the vertices $V = \{S\mid (S,\tau, * \, || \, \mi{start})\in N^*\}$ and 
labeled edges $E = \{(S,i,T)\mid (S,\tau, i \, || \, \bot)\in N^*\text{ and }
(T,\tau::i, * \, || \, \mi{start})\in N^*\}$.

The state graph of some PIDL specification  $\mf{S}$ corresponds to
the semantics of PIDL, i.e., $V=\mc{S}_{\mf{S}}$ and if state $T$ is reachable
from state $S$ in  $G_\mf{S}$, then there is a path from $S$ to $T$.
This justifies confusion of $G_\mf{S}$ and the semantics for $\mf{S}$.

\section{Properties}\label{sec:properties}

In this section, we define properties of rule-based configuration
systems in terms of PIDL and show how they
can be verified with our calculus. In the next section, we present how to
use these properties in order to detect anomalies in a DOPLER
model. We assume a given PIDL specification $\mf{S}$ and its state
graph $G_\mf{S}$ with $V = \{S\mid (S,\tau, * \, || \, \mi{start})\in
N^*\}$ and $E = \{(S,i,T)\mid (S,\tau, i \, || \, \bot)\in N^*\text{
  and } (T,\tau::i, * \, || \, \mi{start})\in N^*\}$.
\begin{itemize}
\item \emph{Inconsistency}: $\mf{S}$ is inconsistent iff there is a
  labeled clause $(S, \tau, * \, || \, \bot) \in \{(S_I, \epsilon, * \,
  || \, \mi{start})\}^*_{\mi{SRInf}}$.
\item \emph{Incompleteness}: Let $\phi$ be a formula over $\Pi$. $\mf{S}$
  is incomplete with respect to $\phi$ iff there is a $S$ in the state
  graph $G_{\mf{S}}$ such that $S$ is rule-terminal and $S \cup \ms{C}
  \not \models \phi$.
\item \emph{Redundancy}: Two rule transitions $\chi_i 
    \leadsto E_i$ and $\chi_j \leadsto E_j \in T_R$ are redundant with 
    respect to a state $T \in V$ iff there are edges $(S, i, T)$ and 
    $(S, j, T) \in E$.
\item \emph{Cycle}: A cycle in $\mf{S}$ is a 
  simple cycle of length greater than one in the state graph
  $G_{\mf{S}}$, i.e., a path of the form $S_1, S_2, \dots,
  S_n$ with $n \geq 3$, $S_1 = S_n$, $(S_i, j_i, S_{i + 1}) \in E$ and
  the vertices $S_2, \dots, S_{n - 1}$ are all different from each
  other.
\item \emph{Confluence}: We distinguish between two types of confluences:
\begin{itemize}
\item $\mf{S}$ is \emph{rule-confluent} iff for each state $S \in V$ the next
rule-terminal state $T \in V$ that can be reached from $S$ is unique.
\item $\mf{S}$ is \emph{user-confluent} iff for states $S, S' \in V$
  that can be reached from $S_I \in V$ via paths $\tau$ and $\tau'$
  respectively, where $\tau$ and $\tau'$ contain the same set of
  indexes that represent user transitions, the next rule-terminal
  state $T \in V$ that can be reached from $S$ and $S'$ is unique.
\end{itemize}
\end{itemize}
Note that these properties are decidable in PIDL because of the
decidability theorem.

\section{A Model of DOPLER}\label{sec:model-dopler}

In this section, we show how PIDL encodes DOPLER models. We first
describe the translation of DOPLER models into the logic. Then, we
describe how anomalies in a DOPLER model can be detected using our PIDL
framework from Section~\ref{sec:pidl}. In the last part of the section,
we evaluate our first prototypical implementation of the PIDL
calculus.

\subsection{Translation}\label{subsec:dopler-to-pidl}

We consider each relevant element of a DOPLER model and explain how it
is represented in a PIDL specification $\mf{S} = (\Pi, S_I, \ms{C},
T_U, T_R)$. We give examples that refer to the DOPLER model
illustrated in Figure~\ref{fig:dopler-example}.

\subsection*{Decisions}
DOPLER decisions are modeled as propositional variables in
$\Pi$. In DOPLER, there are two types of decisions: namely Boolean and 
enumeration decisions.

For each Boolean decision \texttt{d}, we introduce two propositional
variables $d\un\mi{Yes}$ and $d\un\mi{No}$ . This allows us to
distinguish taken from open decisions. If $d\un\mi{Yes}$ is true
then the decision \texttt{d} is assigned to \texttt{true}. If
\texttt{d} is assigned to \texttt{false} then $d\un\mi{No}$ is
true. The following formula represents the fact that \texttt{d} has
not been assigned to a value $\neg d\un\mi{Yes} \land \neg d\un\mi{No}$.

\begin{example}
In the example from
Section~\ref{dec:DOPLER}, the decision \texttt{stainlessSteel}
is represented by the variables $\mi{stainlessSteel}\un\mi{Yes}$ and
$\mi{stainlessSteel}\un\mi{No}$ in PIDL.
\end{example}

In a DOPLER state, a Boolean decision cannot be true and false at the
same time, which has to be considered in the corresponding PIDL
specification as well. One could do this by adding formulas $\neg
(d\un\mi{Yes} \land d\un\mi{No})$ to the constraints $\ms{C}$. An
alternative way is ensuring that this property holds in the initial
state and formulating the transitions so that it is preserved in the
induced states, which is what we did as described below in the
explanations of how we model DOPLER rules, user decisions and the
initial state.

 For each enumeration decision and
each of its options, we introduce a variable denoting that the respective
option is selected.

\begin{example}
 \texttt{casterType} leads to the variables
$\mi{casterType}\un\mi{slab}$,
$\mi{casterType}\un\mi{bloom}$ and
$\mi{casterType}\un\mi{beam}$.
\end{example}

\subsection*{Assets}
For each asset we introduce a propositional variable. If the
variable is set to true in the PIDL model, this corresponds to the
inclusion of the asset in the DOPLER model.

\begin{example}
For example,
$\mi{baleAdapter}$ means the asset \texttt{baleAdapter} is included in
the product.
\end{example}

\subsection*{Visibility condition}
A visibility condition of a decision is modeled as a
propositional formula over $\Pi$.

\begin{example}
 In our DOPLER example (Figure~\ref{fig:dopler-example}),
the visibility condition of \texttt{hydraulicCylinder},
\texttt{containsOnly(casterType, slab) \&\& !taperUnit},
 is encoded by the following formula:

\begin{align*}
&\mi{casterType}\un\mi{slab}\; \land\\ 
&\neg \mi{casterType}\un\mi{bloom}\; \land\\ 
&\neg \mi{casterType}\un\mi{beam}\; \land \\
&\mi{taperUnit}\un\mi{No}. 
\end{align*}
\end{example}

Furthermore, we represent the fact
that a decision is visible as a variable in $\Pi$. For each decision
\texttt{d}, a variable $\mi{Visible}\un{d}$ is introduced.

\begin{example}
The variable $\mi{Visible}\un\mi{stainlessSteel}$ states that
\texttt{stainlesSteel} is visible to the user.
\end{example}

Lastly, for each
decision $d$, the formula 
\begin{equation*}
\phi \rightarrow \mi{Visible}\un\mi{d}
\end{equation*}
is contained in the constraints $\ms{C}$,
 where $\phi$ is the formula
derived from the visibility condition of \texttt{d}. This embodies the
fact that whenever the visibility condition of \texttt{d} is
fulfilled, \texttt{d} is visible.

\begin{example}
 From the DOPLER example (Figure~\ref{fig:dopler-example}),
 we create the following formula denoting if the decision
 \texttt{hydraulicCylinder} is visible to the user:
\begin{align*}
&\mi{casterType}\un\mi{slab} \land \\
 &\neg \mi{casterType}\un\mi{bloom} \land\\
 &\neg \mi{casterType}\un\mi{beam} \land\\
 &\mi{taperUnit}\un\mi{No}\\ 
&\rightarrow \mi{Visible}\un\mi{hydraulicCylinder}.
\end{align*}
\end{example}


\subsection*{Asset inclusion condition}
An asset has an inclusion condition indicating if it is part of
the final product.  It can be translated into a propositional formula over
$\Pi$. 

\begin{example}
For example, the inclusion condition
\texttt{containsOnly(casterType, slab)} of the asset \texttt{baleAdapter} gives
the formula:
\begin{align*}
&\mi{casterType}\un\mi{slab} \land\\
& \neg \mi{casterType}\un\mi{bloom} \land\\
& \neg\mi{casterType}\un\mi{beam}.
\end{align*}
\end{example}

 For each asset \texttt{a}, we derive a
formula $\phi \rightarrow a$, where the inclusion condition of
\texttt{a} is translated into a formula $\phi$ over $\Pi$ that is
added to the constraints $\ms{C}$.

\begin{example}
 To continue the last example, the following formula denotes the respective
inclusion condition:
\begin{align*}
&\mi{casterType}\un\mi{slab} \land\\
& \neg \mi{casterType}\un\mi{bloom} \land\\
& \neg \mi{casterType}\un\mi{beam}\\
& \rightarrow \mi{baleAdapter}.
\end{align*}
\end{example}

 Moreover, for each asset \texttt{a} that
includes another asset \texttt{b}, the formula $a \rightarrow b$ is
contained in $\ms{C}$, where $a$ and $b$ are the propositional
variables expressing the inclusions of assets \texttt{a} and
\texttt{b} respectively. 

Analogously, for each asset \texttt{a} that
excludes an asset \texttt{b}, the formula
 $a \rightarrow \neg b$ is contained in $\ms{C}$.

\begin{example}
 In our example, these are the formulas:
\begin{equation*}
 \mi{baleAdapter}\rightarrow \mi{pCalibthermometer}
\end{equation*}
\begin{equation*}
\mi{calibrator} \rightarrow \neg \mi{pCalibthermometer}.
\end{equation*}
\end{example}

\subsection*{Rules}

For each DOPLER rule which has the form 
\begin{equation*}
\texttt{if <condition> then <action>},
\end{equation*}
we add a rule transition $\chi_i \leadsto E_i$ to $T_R$ as follows:
We translate the \texttt{<condition>} part of a rule
into a formula $\chi_i$ over $\Pi$. The set $E_i$ then contains the
literals that reflect the assignment of values to decisions caused by
the rule's
\texttt{<action>} part.
 If \texttt{true} (\texttt{false}) is
assigned to a Boolean decision \texttt{d} in \texttt{<action>}, then
$E_i$ contains $d\un{Yes}$ ($\neg d\un{Yes}$) and $\neg d\un{No}$
($d\un{No}$). This is to ensure the consistency of the representation
of the Boolean decisions as mentioned above. 
\begin{example}
For example, the rule
\begin{equation*}
\texttt{if !gapChecker then taperUnit = true}
\end{equation*}
becomes the rule transition
\begin{equation*} 
\mi{gapChecker}\un\mi{No} \leadsto \{\mi{taperUnit}\un\mi{Yes}, \neg
\mi{taperUnit}\un\mi{No}\}.
\end{equation*}
\end{example}

 If the decision \texttt{d} is an
enumeration decision that is assigned an option \texttt{o}, then $E_i$
contains $d\un{o}$.

\begin{example}
The rule
\begin{equation*}
\texttt{if molder then setValue(casterType, bloom)}
\end{equation*}
becomes the rule transition
\begin{equation*} 
\mi{molder}\un\mi{Yes} \leadsto \{\mi{casterType}\un\mi{bloom}\}.
\end{equation*}
\end{example}

\subsection*{Decisions taken by the user}
The user transitions $\chi_i \leadsto E_i \in T_U$ model
decision taking by the users in a DOPLER model. In a user
transition $\chi_i \leadsto E_i$, $\chi_i$ states the
conditions that must be fulfilled in order to carry out the user
decision, $E_i$ contains the changes in the set of decisions after the
user taking the decision. For each Boolean decision \texttt{d}, $T_U$
contains two user transitions $\chi_i \leadsto E_i$ and $\chi_{i + 1}
\leadsto E_{i + 1}$. $\chi_i$ and $\chi_{i + 1}$ are the same formula
\begin{equation*}
\mi{Visible}\un{d} \land \neg d\un{Yes} \land \neg d\un{No},
\end{equation*}
stating that \texttt{d} is visible and has not been taken yet.

\begin{example}
Consider the user decision \texttt{stainlessSteel} from the example in
Figure~\ref{fig:dopler-example} that is represented by the following user
transitions $\chi_i \leadsto E_i$ and $\chi_{i + 1} \leadsto E_{i +
  1}$ as follows:
\begin{align*}
\chi_i = \chi_{i + 1} =  &\mi{Visible}\un\mi{stainlessSteel} \land \\
&\neg \mi{stainlessSteel}\un\mi{Yes} \land\\
 &\neg \mi{stainlessSteel}\un\mi{No}.
\end{align*}
$E_i$ and $E_{i+1}$  are sets of literals that denote the update
of the variables after the transition:
\begin{align*} 
E_i &= \{\mi{stainlessSteel}\un{Yes}, \neg \mi{stainlessSteel}\un{No}\}\\
E_{i + 1} &= \{\mi{stainlessSteel}\un{No}, \neg \mi{stainlessSteel}\un{Yes}\}.
\end{align*}
\end{example}

User transitions for enumeration
decisions are analogously obtained.

 In each user transition, we
ensure in $\chi_i$ that the corresponding decision has not
been taken yet. As a consequence, we do not consider user changing
decisions. This does not affect the functionality of the semantics
being discussed because retracting decisions just means reverting to
the state before the decision was taken.

We use rule transitions for DOPLER rule execution and user transitions
for user-decision taking. This is reasonable if we look at how the
rule engine of DOPLER works as described in Section~\ref{dec:DOPLER}:
Once the user has taken a decision, it is checked which rules can be
triggered. Then the action of the rules whose conditions are satisfied
are executed, possibly leading to new checks and executions of
rules. When this procedure is over, the user can take the next
decision. The user cannot take a decision while the rule engine is
operating. We take this into account by considering user transitions
that additionally require a state to be rule-terminal as defined in
Section~\ref{sec:pidl} in order to apply the transition to the state.

\subsection*{Initial state}
The initial state $S_I$ of the PIDL specification of a DOPLER model
consists of all the variables of
$\Pi$ representing the decisions as negative literals. 

The reason why
we have only negative literals here is that we reflect the fact that
in the beginning of the execution of a DOPLER model, nothing is
selected yet, i.e., no value is set for any decision and each
decision has been taken neither by the user nor by any
rule. Also, note that this initial state satisfies the consistency of
Boolean decisions, which is then preserved by the transitions.


\subsection{Detecting DOPLER Anomalies}\label{subsec:pidl-dopler-properties}

By translating a DOPLER model into PIDL we can use our new calculus to
analyze a DOPLER model. We consider the anomalies listed in 
Section~\ref{dec:DOPLER} 
and explain how the calculus detects them. In the following, we assume
a DOPLER model and its corresponding PIDL specification $\mf{S}$ with
its state graph $S_{\mf{S}}$ as defined in Section~\ref{sec:pidl}.

\begin{itemize}
\item \textbf{Inconsistency:} Consistency properties of the DOPLER
  model are modeled as formulas in $\ms{C}$. Then inconsistency of the
  DOPLER model corresponds to inconsistency of $\mf{S}$. As one
  example of such a property, an enumeration decision $d$ has a
  minimum number and a maximum number of options that can be
  selected. With the variables of $\Pi$, propositional formulas $\phi$
  stating these values restrictions can be derived. These $\phi$ are
  then contained in $\ms{C}$. 
\begin{example}
The formula 
\begin{align*}
&\neg (\mi{casterType}\un\mi{slab} \land\\
 &\mi{casterType}\un\mi{bloom} \land\\
& \mi{casterType}\un\mi{beam})
\end{align*}
 says that \texttt{casterType} cannot have all three values selected
 at the same time.
\end{example}

\item \textbf{Incompleteness:}  The DOPLER incompleteness test case
 is expressed as a formula $\phi$ over $\Pi$ as the following example depicts.
\begin{example}
 Consider the incompleteness test case from the DOPLER example in
Figure~\ref{fig:dopler-example}: The modeler expects the value of
\texttt{hydraulicCylinder} to be set automatically, after
\texttt{stainlessSteel} is assigned a value.

This is expressed as the following formula
\begin{align*}
\phi = & \mi{stainlessSteel}\un\mi{Yes} \lor\\
 &\mi{stainlessSteel}\un\mi{No}\\
 & \rightarrow \\
&(\mi{casterType}\un\mi{slab} \lor \\
  &\mi{casterType}\un\mi{bloom} \lor\\
& \mi{casterType}\un\mi{beam}) \land \\
 & (\mi{hydraulicCylinder}\un\mi{Yes} \lor \\ 
 & \mi{hydraulicCylinder}\un\mi{No}).
\end{align*}
 Then it is checked if $\mf{S}$
  is incomplete with respect to $\phi$.
\end{example}

\item \textbf{Redundancy:} Two DOPLER rules are redundant iff there is
  a state $S \in V$ such that the two rule transitions that represent
  these rules are redundant with respect to $S$.

\item \textbf{Cyclicity:} A cycle in the DOPLER model is detected by
  checking if $\mf{S}$ has a cycle.

\item \textbf{Confluence:} We have confluence in the DOPLER model iff
  $\mf{S}$ is rule-confluent and user-confluent.

\item \textbf{Asset Inclusion Conflicts:} As mentioned before, one way
  to model the inclusion of assets in products by inclusion conditions
  and includes- and excludes-relationships between assets is to
  translate them into formulas over $\Pi$ that are contained in the
  constraints $\ms{C}$. Assets conflicts can then be identified by
  checking inconsistency of $\mf{S}$.
\end{itemize}

\subsection{Implementation}

We made a first implementation of the PIDL framework to see how it could be
used in practice. The tool takes DOPLER models as inputs and checks them
for anomalies. It translates a DOPLER model to 
elements of PIDL, creating a specification as described in the previous 
subsection. A state is the current truth assignment of the 
variables corresponding to the DOPLER decisions. Inconsistency of the states
and transitions to new states are then determined by superposition-based 
SAT solving, following the calculus in Section~\ref{sec:pidl}. The state graph
of the specification is produced, which is used to detect graph-based 
properties such as cyclicity by standard graph algorithms. Our first prototypical 
implementation does currently not contain the confluence check.

Table~\ref{tab:implementation} shows the results of running our implementation
on the example in Figure~\ref{fig:dopler-example}, displaying what kind of
anomalies were found. All in all, 99 states were created
during the run, which took $0.037$ seconds on an
 Intel Xeon E5-4640 running at 2.4~GHz and 512~GB of RAM. 
The program detected 12 inconsistent states. Incompleteness was found for 
7~states. Out of the 99~states, 10~states showed rule redundancy and in 12 cases there were
conflicts in the asset inclusions. The mentioned cycle in the DOPLER
model example was identified.

Additionally, we ran the implementation on a set of randomly generated 
DOPLER models. We used models with 20, 60 and 100 Boolean
decision variables respectively. Each model contains a set of  
random rules according to the predefined fixed form 
\texttt{if (d||[!]e\&\&[!]f) then g = [true/false]}, where \texttt{d}, 
\texttt{e}
\texttt{f} and \texttt{g} are pairwise distinct Boolean decision variables, 
with \texttt{e} and \texttt{f} possibly being negated. The number of the rules
are such that we have a ratio of 1:1.5 between variables and rules. The 
visibility of decisions is organized such that at most one half of the variables
are visible, but visible variables are not allowed to appear on the action
sides of the rules. This is to ensure that the rule's contribution to the 
states generation is not diminished. Consequently, each generated model may 
differ in the number of visible decisions. We furthermore added random  
constraint clauses of the form \texttt{([!]d||[!]e||![f])} to get more 
realistic examples. 
Without any constraints,
the models would amount to a mere enumeration of reachable states. We used a
ratio of 1:1 between variables and constraints. For each of the three
model sizes, we generated 20 instances. 

The results are shown in 
Table~\ref{tab:implementation2}. If no 
inconsistency with respect to the
random constraints can be found, a triple is shown indicating how many
states were generated,  if a cycle was detected (Y) or not (N) and the 
number of redundant rules applications. For example, the result 211/N/8 of the 
model rnd\un10 means that a state graph with 211 states was created, there was 
no cycle and there were 8 cases of redundant rules applications. Consistent 
models only occurred with 20 variables. Although most
models are still solved in a relatively short time, one can see that problems 
get
harder with rising numbers of variables. In the group of models with 60 
variables instances that required up to several minutes run-time can be found,
whereas most of the examples with 20 variables stayed under one second. 
Finally, when dealing with 100 variables, we see four cases in the table 
where the run
was aborted by the system after approximately 12 minutes.

The potential search 
space has $3^{v + a}$ states, where $v$ is the number of visible decisions and
$a$ is the number of decisions occurring on the action sides of the rules 
(in our experiments mostly $a = n - v$ with $n$ being the number of variables).
Improvements to this first implementation can reduce the search space. This
could be done by taking invariants among the states into account and by 
considering similarities and dependencies between them. Nevertheless, it can 
be seen that PIDL can in fact be turned into a useful software system for 
the practical analysis of rule-based systems.

\begin{table}
  \caption{Anomalies in the DOPLER model example.}
  \label{tab:implementation}
  \centering
  \begin{tabular}{|c|c|c|}
    \hline 
    & Number of States \\ 
    \hline 
    \hline 
    Total & 99 \\ 
    \hline
    Inconsistency & 12\\ 
    \hline 
    Incompleteness & 7 \\ 
    \hline
    Redundancy & 10 \\ 
    \hline 
    Cycle & *detected*\\ 
    \hline 
    Asset Inclusion Conflicts & 12 \\ \hline
  \end{tabular} 
\end{table}

\begin{table}
  \caption{Generated random DOPLER models.}
  \label{tab:implementation2}
  \centering
  \begin{tabular}{|c|c|c|c|}
    \hline 
    \multicolumn{4}{|c|} {20 variables, 30 rules}\\
    \hline
    Name & Visible Variables & Time & Results\\ 
    \hline
    \hline
    rnd\un1 & 5 & 0m0.05s & inconsistent \\
    \hline
    rnd\un2 &  3 & 0m1.00s & 1079/Y/0\\
    \hline
    rnd\un3 & 6 &  0m1.45s & inconsistent\\
    \hline  
    rnd\un4 & 4 & 0m0.07s & inconsistent \\
    \hline
    rnd\un5 & 4 &  0m0.05s & inconsistent\\
    \hline
    rnd\un6 & 6 & 0m0.04s & inconsistent\\
    \hline
    rnd\un7 & 3 &  0m0.03s & inconsistent\\
    \hline
    rnd\un8 & 2 &  0m0.05s & inconsistent \\
    \hline
    rnd\un9 & 3 &  0m0.09s & inconsistent \\
    \hline
    rnd\un10 & 2 & 0m0.26s & 211/N/8 \\
    \hline
    rnd\un11 & 3 &  0m0.04s & inconsistent \\
    \hline
    rnd\un12 & 2 &  0m0.04s & 9/N/0\\
    \hline
    rnd\un13 & 5 &  0m0.20s & inconsistent \\
    \hline
    rnd\un14 & 7 &  0m0.06s & inconsistent \\
    \hline
    rnd\un15 & 3 &  0m0.02s & inconsistent\\
    \hline
    rnd\un16 & 5 &  0m1.11s & inconsistent\\
    \hline
    rnd\un17 & 3 &  0m0.11s & inconsistent\\
    \hline
    rnd\un18 & 4 &  0m0.24s & inconsistent \\
    \hline
    rnd\un19 & 3 &  0m0.61s & 558/Y/240 \\
    \hline
    rnd\un20 & 4 &  0m0.43s & inconsistent\\
    \hline
    \hline 

    \multicolumn{4}{|c|} {60 variables, 90 rules}\\
    \hline
    Name & Visible Variables & Time & Results\\ 
    \hline
    \hline
    rnd\un21 & 16 & 0m0.47s & inconsistent \\
    \hline
    rnd\un22 & 10 & 0m0.61s & inconsistent \\
    \hline
    rnd\un23 & 11 & 0m4.38s & inconsistent \\
    \hline
    rnd\un24 & 12 & 0m2.84s & inconsistent \\
    \hline
    rnd\un25 & 13 & 7m44.81s & inconsistent \\
    \hline
    rnd\un26 & 14 & 4m51.23s & inconsistent\\
    \hline
    rnd\un27 & 14 & 0m0.38s & inconsistent \\
    \hline
    rnd\un28 & 13 & 0m0.39s & inconsistent \\
    \hline
    rnd\un29 & 15 & 0m0.51s & inconsistent \\
    \hline
    rnd\un30 & 15 & 0m0.77s & inconsistent \\
    \hline
    rnd\un31 & 15 & 0m0.70s & inconsistent \\
    \hline
    rnd\un32 & 14 & 0m0.30s & inconsistent \\
    \hline
    rnd\un33 & 11 & 0m1.01s & inconsistent \\
    \hline
    rnd\un34 & 15 & 0m0.65s & inconsistent \\
    \hline
    rnd\un35 & 12 & 0m36.00s & inconsistent \\
    \hline
    rnd\un36 & 16 & 0m0.50s & inconsistent \\
    \hline
    rnd\un37 & 9 & 0m2.00s & inconsistent \\
    \hline
    rnd\un38 & 14 & 0m0.40s & inconsistent \\
    \hline
    rnd\un39 & 11 & 2m13.94s& inconsistent \\
    \hline
    rnd\un40 & 10 & 0m44.69s & inconsistent \\
    \hline
    \hline
    \multicolumn{4}{|c|} {100 variables, 150 rules}\\
    \hline
    Name & Visible Variables & Time & Results\\ 
    \hline
    \hline
    rnd\un41 & 24 & 0m1.55s & inconsistent\\
    \hline
    rnd\un42 & 24 & >12m & - \\
    \hline
    rnd\un43 & 25 & 0m2.42s & inconsistent \\
    \hline
    rnd\un44 & 22 & >12m & - \\
    \hline
    rnd\un45 & 18 & 5m59.85s & inconsistent \\
    \hline
    rnd\un46 & 29 & 0m0.81s & inconsistent \\
    \hline
    rnd\un47 & 20 & 0m1.51s & inconsistent \\
    \hline
    rnd\un48 & 22 & 0m2.84s & inconsistent \\
    \hline
    rnd\un49 & 23 & 7m15.73s & inconsistent \\
    \hline
    rnd\un50 & 19 & 0m42.68s & inconsistent \\
    \hline
    rnd\un51 & 26 & >12m & -\\
    \hline
    rnd\un52 & 28 & 0m16.12s & inconsistent\\
    \hline
    rnd\un53 & 21 & 0m1.28s& inconsistent\\
    \hline
    rnd\un54 & 17 & 0m0.73s & inconsistent\\
    \hline
    rnd\un55 & 18 & 0m1.48s & inconsistent \\
    \hline
    rnd\un56 & 25 & 0m2.18s & inconsistent \\
    \hline 
    rnd\un57 & 20 & 0m1.34s & inconsistent\\
    \hline 
    rnd\un58 & 21 & >12m & -\\
    \hline
    rnd\un59 & 21 & 0m1.13s  & inconsistent \\
    \hline
    rnd\un60 & 23 & 0m1.56s  & inconsistent \\
    \hline   
    
  \end{tabular} 
\end{table}

\section{Related Work}\label{dec:relatedwork}

Verification of configuration knowledge bases has been tackled by many
researchers on varying levels of details and granularity. Yang et
al.~\cite{Yang_2003} present an approach based on petri nets, where
all rules are first normalized into Horn clauses and transformed to
petri nets.

Verification of models for product line engineering (typically feature
models) have also been intensively studied in literature. Some
approaches verify development artifacts \cite{Lauenroth_2009} and some
others verify that the variability specified by a feature model is
correctly implemented in code~\cite{Czarnecki_2006}. Verification of
the models themselves has been studied by Post and Sinz
\cite{Post_2008}, where the authors describe the variants of the
product line using a meta-program. All these approaches follow a
constraint-based approach.  Our approach deals with rules, which are
easy to specify for the modelers but rather complex to verify and
maintain.

Logical representation of feature models has been previously discussed by
Czarnecki et al. \cite{Czarnecki_2007}. Other analysis techniques
available for product line models include approaches based on SAT
solvers \cite{Mendonca_2009}, atomic sets \cite{Segura08},
BDDs \cite{Zhang_2008} and CSPs \cite{Benavides2010615} etc.  The
primary difference between all these contributions and our work is
that these approaches do not consider the interactive nature of the
configuration process and the rule-based specification (as opposed to
constraints) of restrictions on the models.

In the context of propositional logic there are various extensions to
propositional logic with time~\cite{firstP77} or dynamic propositional
logic~\cite{FischerLadner79} also based on superposition~\cite{SudaWeidenbach12,SudaWeidenbach12lpar}.  
For these logics there exists a variety
of modern proof calculi. However, they do not directly support our transition
semantics via language constructs. Nevertheless, the implementation
techniques used for these logics have also potential for improving
our current prototypical implementation.

Specific to our approach is the support of rules of the from $A \land
\phi \leadsto \{ \neg A,\ldots\}$ enabling revision of a
decision. Such rules cannot be modeled in many of the aforementioned
approaches or would lead to an inconsistency. Unique to PIDL is the
concept of rule-terminal states that are a prerequisite for some rules
(in case of DOPLER user decisions) to be applied.

\section{Conclusions}\label{sec:conclusions}

In this paper, we have defined the new logic PIDL that provides
detailed models for rule-based configuration systems.  In particular,
it supports decision revision as expressed by rules of the form $A
\land \phi \leadsto \{ \neg A,\ldots\}$ and the concept of
rule-terminal states.  In addition, we provide a sound and complete
calculus for PIDL that is based on the ideas of superposition. This
calculus constitutes a decision procedure that analyzes the following
properties of rule-based systems: inconsistency, incompleteness,
redundancy, absence of cycles, confluence and conflicts of asset
inclusion.

We have presented the automatic translation of DOPLER models to PIDL.
DOPLER is a rule-based configuration system currently in use at
Siemens.  Furthermore, we showed by a first prototypical
implementation that PIDL can in fact be turned into a useful software
system for the practical analysis of rule-based systems.


\section*{Acknowledgment}
This work was partly supported by Siemens. We would like to 
thank Martin Suda for his helpful advice and fruitful discussions.

\bibliographystyle{IEEEtran}

\bibliography{paper}

\begin{thebibliography}{10}
\providecommand{\url}[1]{#1}
\csname url@samestyle\endcsname
\providecommand{\newblock}{\relax}
\providecommand{\bibinfo}[2]{#2}
\providecommand{\BIBentrySTDinterwordspacing}{\spaceskip=0pt\relax}
\providecommand{\BIBentryALTinterwordstretchfactor}{4}
\providecommand{\BIBentryALTinterwordspacing}{\spaceskip=\fontdimen2\font plus
\BIBentryALTinterwordstretchfactor\fontdimen3\font minus
  \fontdimen4\font\relax}
\providecommand{\BIBforeignlanguage}[2]{{%
\expandafter\ifx\csname l@#1\endcsname\relax
\typeout{** WARNING: IEEEtran.bst: No hyphenation pattern has been}%
\typeout{** loaded for the language `#1'. Using the pattern for}%
\typeout{** the default language instead.}%
\else
\language=\csname l@#1\endcsname
\fi
#2}}
\providecommand{\BIBdecl}{\relax}
\BIBdecl

\bibitem{DhunganaGR11}
D.~Dhungana, P.~Gr{\"u}nbacher, and R.~Rabiser, ``The {DOPLER} meta-tool for
  decision-oriented variability modeling: a multiple case study,'' \emph{Autom.
  Softw. Eng.}, vol.~18, no.~1, pp. 77--114, 2011.

\bibitem{DhunganaHR10}
D.~Dhungana, P.~Heymans, and R.~Rabiser, ``A formal semantics for
  decision-oriented variability modeling with {DOPLER},'' in \emph{Fourth
  International Workshop on Variability Modelling of Software-Intensive
  Systems, Linz, Austria, January 27-29, 2010. Proceedings}, ser. ICB-Research
  Report, D.~Benavides, D.~S. Batory, and P.~Gr{\"u}nbacher, Eds.,
  vol.~37.\hskip 1em plus 0.5em minus 0.4em\relax Universit\"at Duisburg-Essen,
  2010, pp. 29--35.

\bibitem{BachmairGanzinger01handbook}
L.~Bachmair and H.~Ganzinger, ``Resolution theorem proving,'' in \emph{Handbook
  of Automated Reasoning}, A.~Robinson and A.~Voronkov, Eds.\hskip 1em plus
  0.5em minus 0.4em\relax Elsevier, 2001, vol.~I, ch.~2, pp. 19--99.

\bibitem{NieuwenhuisRubio01handbook}
R.~Nieuwenhuis and A.~Rubio, ``Paramodulation-based theorem proving,'' in
  \emph{Handbook of Automated Reasoning}, A.~Robinson and A.~Voronkov,
  Eds.\hskip 1em plus 0.5em minus 0.4em\relax Elsevier, 2001, vol.~I, ch.~7,
  pp. 371--443.

\bibitem{Weidenbach01handbook}
C.~Weidenbach, ``Combining superposition, sorts and splitting,'' in
  \emph{Handbook of Automated Reasoning}, A.~Robinson and A.~Voronkov,
  Eds.\hskip 1em plus 0.5em minus 0.4em\relax Elsevier, 2001, vol.~2, ch.~27,
  pp. 1965--2012.

\bibitem{Yang_2003}
S.~J.~H. Yang, J.~J.~P. Tsai, and C.-C. Chen, ``Fuzzy rule base systems
  verification using high-level petri nets,'' \emph{IEEE Trans. Knowl. Data
  Eng.}, vol.~15, no.~2, pp. 457--473, Feb. 2003.

\bibitem{Lauenroth_2009}
K.~Lauenroth, K.~Pohl, and S.~Toehning, ``Model checking of domain artifacts in
  product line engineering,'' in \emph{ASE 2009, 24th IEEE/ACM International
  Conference on Automated Software Engineering, Auckland, New Zealand, November
  16-20, 2009}.\hskip 1em plus 0.5em minus 0.4em\relax IEEE Computer Society,
  2009, pp. 269--280.

\bibitem{Czarnecki_2006}
K.~Czarnecki and K.~Pietroszek, ``Verifying feature-based model templates
  against well-formedness {OCL} constraints,'' in \emph{Generative Programming
  and Component Engineering, 5th International Conference, GPCE 2006, Portland,
  Oregon, USA, October 22-26, 2006, Proceedings}, S.~Jarzabek, D.~C. Schmidt,
  and T.~L. Veldhuizen, Eds.\hskip 1em plus 0.5em minus 0.4em\relax ACM, 2006,
  pp. 211--220.

\bibitem{Post_2008}
H.~Post and C.~Sinz, ``Configuration lifting: Verification meets software
  configuration,'' in \emph{23rd IEEE/ACM International Conference on Automated
  Software Engineering (ASE 2008), 15-19 September 2008, L'Aquila,
  Italy}.\hskip 1em plus 0.5em minus 0.4em\relax IEEE, 2008, pp. 347--350.

\bibitem{Czarnecki_2007}
K.~Czarnecki and A.~Wasowski, ``Feature diagrams and logics: There and back
  again,'' in \emph{Software Product Lines, 11th International Conference, SPLC
  2007, Kyoto, Japan, September 10-14, 2007, Proceedings}.\hskip 1em plus 0.5em
  minus 0.4em\relax IEEE Computer Society, 2007, pp. 23--34.

\bibitem{Mendonca_2009}
M.~Mendon\c{c}a, A.~Wasowski, and K.~Czarnecki, ``{SAT}-based analysis of
  feature models is easy,'' in \emph{Software Product Lines, 13th International
  Conference, SPLC 2009, San Francisco, California, USA, August 24-28, 2009,
  Proceedings}, ser. ACM International Conference Proceeding Series, D.~Muthig
  and J.~D. McGregor, Eds., vol. 446.\hskip 1em plus 0.5em minus 0.4em\relax
  ACM, 2009, pp. 231--240.

\bibitem{Segura08}
S.~Segura, ``Automated analysis of feature models using atomic sets,'' in
  \emph{Software Product Lines, 12th International Conference, SPLC 2008,
  Limerick, Ireland, September 8-12, 2008, Proceedings. Second Volume
  (Workshops)}, S.~Thiel and K.~Pohl, Eds.\hskip 1em plus 0.5em minus
  0.4em\relax Lero Int. Science Centre, University of Limerick, Ireland, 2008,
  pp. 201--207.

\bibitem{Zhang_2008}
W.~Zhang, H.~Yan, H.~Zhao, and Z.~Jin, ``A {BDD}-based approach to verifying
  clone-enabled feature models' constraints and customization,'' in \emph{High
  Confidence Software Reuse in Large Systems, 10th International Conference on
  Software Reuse, ICSR 2008, Beijing, China, May 25-29, 2008, Proceedings},
  ser. Lecture Notes in Computer Science, H.~Mei, Ed., vol. 5030.\hskip 1em
  plus 0.5em minus 0.4em\relax Springer, 2008, pp. 186--199.

\bibitem{Benavides2010615}
D.~Benavides, S.~Segura, and A.~R. Cort{\'e}s, ``Automated analysis of feature
  models 20 years later: A literature review,'' \emph{Inf. Syst.}, vol.~35,
  no.~6, pp. 615 -- 636, 2010.

\bibitem{firstP77}
A.~Pnueli, ``The temporal logic of programs,'' in \emph{18th Annual Symposium
  on Foundations of Computer Science, Providence, Rhode Island, USA, 31 October
  - 1 November 1977}.\hskip 1em plus 0.5em minus 0.4em\relax IEEE Computer
  Society, 1977, pp. 46--57.

\bibitem{FischerLadner79}
M.~J. Fischer and R.~E. Ladner, ``Propositional dynamic logic of regular
  programs,'' \emph{J. Comput. Syst. Sci.}, vol.~18, no.~2, pp. 194--211, 1979.

\bibitem{SudaWeidenbach12}
M.~Suda and C.~Weidenbach, ``A {PLTL}-prover based on labelled superposition
  with partial model guidance,'' in \emph{Automated Reasoning - 6th
  International Joint Conference, IJCAR 2012, Manchester, UK, June 26-29, 2012.
  Proceedings}, ser. Lecture Notes in Computer Science, B.~Gramlich, D.~Miller,
  and U.~Sattler, Eds., vol. 7364.\hskip 1em plus 0.5em minus 0.4em\relax
  Springer, 2012, pp. 537--543.

\bibitem{SudaWeidenbach12lpar}
------, ``Labelled superposition for {PLTL},'' in \emph{Logic for Programming,
  Artificial Intelligence, and Reasoning - 18th International Conference,
  LPAR-18, M{\'e}rida, Venezuela, March 11-15, 2012. Proceedings}, ser. Lecture
  Notes in Computer Science, N.~Bj{\o}rner and A.~Voronkov, Eds., vol.
  7180.\hskip 1em plus 0.5em minus 0.4em\relax Springer, 2012, pp. 391--405.

\end{thebibliography}

\end{document}